\documentclass[iop]{emulateapj} 
\usepackage{natbib,graphicx,amsmath,amsthm,ulem,color}
\usepackage{threeparttable}

\shorttitle{Stellar Companion to HD~8673 }
\shortauthors{Roberts et al.}

\newcommand{\be}{\begin{equation}}
\newcommand{\ee}{\end{equation}}

\def\deg{^\circ}

\begin{document}

\title{Know the Star, Know the Planet. IV. A Stellar Companion to the Host Star of the Eccentric Exoplanet HD~8673b}

\author{Lewis C. Roberts, Jr.\altaffilmark{1},  
Brian D. Mason\altaffilmark{2}, 
Christopher R. Neyman\altaffilmark{3},
Yanqin Wu\altaffilmark{4}, 
Reed L. Riddle\altaffilmark{5}, 
J. Christopher Shelton\altaffilmark{1}, 
John Angione\altaffilmark{1,6}, 
Christoph Baranec\altaffilmark{7}, 
Antonin Bouchez\altaffilmark{8,9}, 
Khanh Bui\altaffilmark{5}, 
Rick Burruss\altaffilmark{1}, 
Mahesh Burse\altaffilmark{10}, 
Pravin Chordia\altaffilmark{10},
Ernest Croner\altaffilmark{5},  
Hillol Das\altaffilmark{10},
Richard G. Dekany\altaffilmark{5}, 
Stephen Guiwits\altaffilmark{11}, 
David Hale\altaffilmark{5}, 
John Henning\altaffilmark{5}, 
Shrinivas Kulkarni\altaffilmark{5}
Nicholas Law\altaffilmark{12}, 
Dan McKenna\altaffilmark{5},
Jennifer Milburn\altaffilmark{5}, 
Dean Palmer\altaffilmark{1},  
Sujit Punnadi\altaffilmark{10}, 
A.N. Ramaprakash\altaffilmark{10}, 
Jennifer E. Roberts\altaffilmark{1}, 
Shriharsh P. Tendulkar\altaffilmark{5},  
Thang Trinh\altaffilmark{1}, 
Mitchell Troy\altaffilmark{1}, 
Tuan Truong\altaffilmark{1}, 
Jeff Zolkower\altaffilmark{5}, 
}

\altaffiltext{1}{Jet Propulsion Laboratory, California Institute of Technology, 4800 Oak Grove Drive, Pasadena CA 91109, USA}\email{lewis.c.roberts@jpl.nasa.gov}
\altaffiltext{2}{U.S. Naval Observatory, 3450 Massachusetts Avenue, NW, Washington, DC 20392-5420, USA}
\altaffiltext{3}{W. M. Keck Observatory, California Association for Research in Astronomy, 65-1120 Mamalahoa Hwy, Kamuela, HI 96743}
\altaffiltext{4}{Department of Astronomy and Astrophysics, University of Toronto, ON M5S 3H4, Canada}
\altaffiltext{5}{Division of Physics, Mathematics, and Astronomy, California Institute of Technology, Pasadena, CA 91125, USA}
\altaffiltext{6}{Deceased}
\altaffiltext{7}{Institute for Astronomy, University of Hawai\textquoteleft i at M\={a}noa, Hilo, HI 96720-2700, USA}
\altaffiltext{8}{GMTO Corp., 251 S. Lake Ave., Pasadena, CA 91101, USA}
\altaffiltext{9}{Observatories of the Carnegie Institution for Science, 813 Santa Barbara St., Pasadena, CA 91101, USA}
\altaffiltext{10}{Inter-University Centre for Astronomy \& Astrophysics, Ganeshkhind, Pune, 411007, India} 
\altaffiltext{11}{Seismological Laboratory, California Institute of Technology, 1200 E. California Blvd., Pasadena, CA 91101, USA}
\altaffiltext{12}{Department of Physics and Astronomy, University of North Carolina at Chapel Hill, Chapel Hill, NC 27599-3255, USA}


\begin{abstract}

HD~8673 hosts a massive exoplanet in a highly eccentric orbit (e=0.723). Based on two epochs of speckle interferometry a previous publication identified a candidate stellar companion.  We observed HD~8673 multiple times with the 10 m Keck II telescope, the 5 m Hale telescope, the 3.63 m AEOS telescope and the 1.5m Palomar telescope in a variety of filters with the aim of confirming and characterizing the stellar companion.  We did not detect the candidate companion, which we now conclude was a false detection, but we did detect a fainter companion.  We collected astrometry and photometry of the companion on six epochs in a variety of filters.  The measured differential photometry enabled us to determine that the companion is an early M dwarf with a mass estimate of 0.33-0.45 M$_\odot$. The companion has a projected separation of 10 AU, which is one of the smallest projected separations of an exoplanet host binary system.  Based on the limited astrometry collected, we are able to constrain the orbit of the stellar companion to a semi-major axis of 35--60 AU,  an eccentricity $\leq$ 0.5 and an inclination of 75--85$^\circ$.  The stellar companion has likely strongly influenced the orbit of the exoplanet and quite possibly explains its high eccentricity.
\end{abstract}



\section{INTRODUCTION}

HD~8673 (HR~410 = HIP~6702 = WDS~01262+3435) is identified as a member of the larger Hyades Moving Group although it does not fit the Hyades Li-temperature pattern or the Hyades Fe abundances \citep{boesgaard1988}. There is a spread in the age estimates of the star. \citet{saffe2005} derived an age of $2.8^{+0.5}_{-0.7}$ Gyr from isochrones and an age  of 8.7 Gyr from Fe/H abundances.   \citet{holmberg2009} produced a similar age of 2.5 Gyr using the photometric technique.  Using isochrones, \citet{valenti2005}, computed an age of 4.29 Gyr. This variation in age estimates is normal due to a number of factors \citep{saffe2005} and HD~8673 is probably somewhere in age between the Hyades and the Sun. It can be broadly described as a nearby solar type star. It was spectroscopically classified  as an F7V  \citep{boesgaard1990} and has a distance of 36.1$\pm$0.5 pc \citep{vanLeeuwen2007}. \citet{tsantaki2014} computed that it ad a T$_{eff}$ of 6472$\pm$64, a mass of 1.56$\pm$0.10 M$_\odot$ and a radius of 1.23$\pm$0.07 R$_\odot$.

A substellar companion orbiting HD~8673 was detected via radial velocity (RV) measurements by \citet{hartmann2010}. HD~8673b has a minimum mass of 14.2$\pm$1.6 $M_{Jup}$, suggested that it is probably  a low-mass brown dwarf rather than a planet.  The companion is in a high eccentricity (e=0.723$\pm$0.016) orbit with a period of  1634$\pm$17 days. The orbit has a semi-major axis of 3.02$\pm$0.15 AU. The detection was part of an effort to detect sub-stellar companions to F-type stars in order to increase the statistics of exoplanets around stars more massive than the Sun.   

In an earlier paper, we  described a candidate stellar companion to HD~8673 \citep{mason2011}. In order to confirm the candidate companion and determine the impact of the stellar companion on the orbital dynamics of the planetary system, we conducted a series of adaptive optics (AO) observations using four different telescopes. Those observations now lead us to conclude that our previous detection was spurious.  However, we detected another companion  that appears to be a bound low-mass stellar companion in a century long orbit. Details of the observations are in Section \ref{observations}, and the analysis of the companion is detailed in Section \ref{analysis}. Finally we discuss the system in Section \ref{discussion}.

\section{OBSERVATIONS}\label{observations}

We observed HD~8673 with four telescopes, each equipped with AO.  Two of the AO systems operate in the visible, while two operate in the near-IR.  

\subsection{KECK II OBSERVATIONS}\label{keck}

We used the Keck II telescope, its AO system \citep{wizinowich2000}, and the NIRC2 instrument to observe the system at three epochs in 2011 and 2012.  Each time, we collected data in multiple filters with multiple coadds. Since the existence of the companion was still questionable, the multiple filters allowed us to determine if the object was an astronomical object or a quasi-static speckle whose position is wavelength dependent.

After reducing each image, we measured the astrometry and photometry using the \textit{fitstars} algorithm on each image  \citep{tenbrummelaar1996, tenbrummelaar2000}.  Photometric error bars were set equal to the standard deviation of the measurements from all the images. We computed the weighted mean of the astrometry from the images taken in different filters on the same night, with the weight being the number of coadds for each image. The error bar was set to the standard deviation of the results. That covers much of the random errors, but does not cover the systematic errors. We expect those to be smaller than those of the Advanced Electro-Optical System (AEOS) (Section \ref{aeos}) or Hale data  (Section \ref{hale}) because of the smaller point spread function (PSF) arising from the larger aperture size, but we do not have a way of quantifying those errors.   The astrometry is given in Table \ref{astrometry_table}; it lists the Besselian date of the observations, the position angle and separation of the companion relative to the primary and the telescope used to collect the measurements. The photometric measurements are listed in Table \ref{photometry_table}.  The table lists the measured magnitude difference in each filter, the central wavelength of those filters, and the telescope used to take the observations.
 
\begin{deluxetable}{cccl}
\tablewidth{0pt}
\tablecaption{Astrometry\label{astrometry_table}}
\tablehead{\colhead{UT} & \colhead{$\theta$ (\degr)} & \colhead{$\rho$ (\arcsec) } & \colhead{Telescope}}
 \startdata
 2004.7871 & 302.3$\pm$1.0\phn    & 0.31\phn$\pm$0.02\phn    &   AEOS \\
 2011.5366 & 332.3$\pm$1.0\phn    & 0.310$\pm$0.005         & Keck II\\

 2011.6402 & 329.5$\pm$1.0\phn    &    0.308$\pm$0.1\phn\phn    & Hale \\
 2011.8671 & 333.4$\pm$0.57       & 0.308$\pm$0.003       & Keck II \\
 2012.4837 & 335.2$\pm$0.63       & 0.308$\pm$0.003       & Keck II \\
 2013.7426 & 339.3$\pm$1.68       & $0.32\phn\pm$0.02\phn &  Hale 
\enddata
\end{deluxetable}

\begin{figure}[htb]
  \begin{center}
\centerline{\includegraphics[width=0.49\textwidth,trim=80 200 80 200,clip]{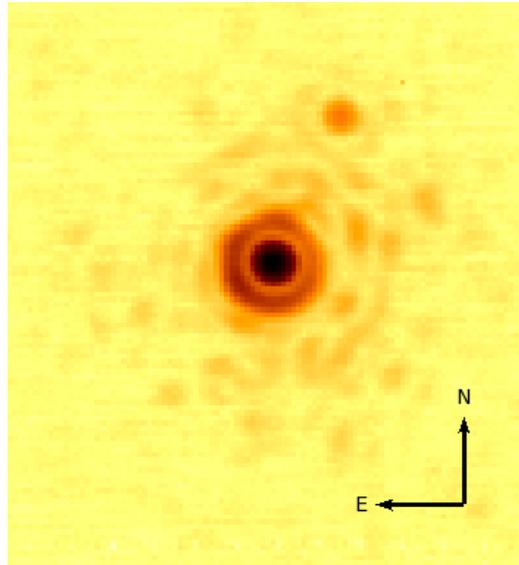}}
  \end{center}
   \caption{$Br\gamma$ image of the HD~8673 binary system taken with NIRC2 at the Keck II telescope on 12 June 2012. North is up in the image and East is to the left.  This is a sub-image of the full image and is approximately 1\farcs2  across. }
\label{keck_image}
\end{figure}

\subsection{HALE  OBSERVATIONS}\label{hale}

\begin{figure}[htb]
  \begin{center}
\centerline{\includegraphics[width=0.49\textwidth,trim=100 280 120 200,clip]{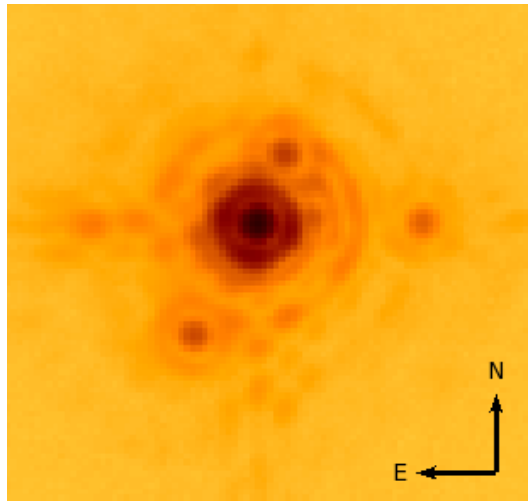}}
  \end{center}
 \caption{$Ks$ image of the HD~8673 binary system taken with PHARO an
``d the PALM 3000 AO system at the Hale telescope on 2013 September 28 UT.  The companion is the object to the upper right.  There is a ghost image caused by the neutral density filter to the lower left.  There is a large speckle to the right of the primary and a mirror reflection of that speckle to the left of the primary.  The image has the same orientation as Fig. \ref{keck_image}.  This is a sub-image of the full 25'' field of view and is approximately 2\farcs5 across.
}
 \label{pharo_image}
\end{figure}

We observed HD~8673 on 2011 August 22 UT and 2013 September 28 UT with the Palomar Observatory Hale 5 m telescope using the PALM 3000 AO system \citep{dekany2013} and the PHARO near-IR camera \citep{hayward2001}.   HD~8673 was centered in the detector's 25\arcsec~field of view.  In 2011, we collected 10 frames of the star in $Br\gamma$ filter and after reducing the data, the frames were coadded.  In 2013, we collected 50 frames in the $K_{s}$ filter.  After the individual frames were calibrated, we created five images by coadding 10 frames into each image, allowing us to analyze the precision of the measurements. The \textit{fitstars} algorithm was used to measure the astrometry and photometry of the objects.  Photometric error bars were assigned using the technique described in \citet{roberts2005}.  The resulting astrometry and photometry are listed in Tables \ref{astrometry_table} and \ref{photometry_table}. Figure \ref{pharo_image} shows an image from the 2013 data collection.

\subsection{AEOS OBSERVATIONS}\label{aeos}
 
\begin{figure}[htb]
  \begin{center}
\centerline{\includegraphics[width=0.49\textwidth,trim=100 200 50 180,clip]{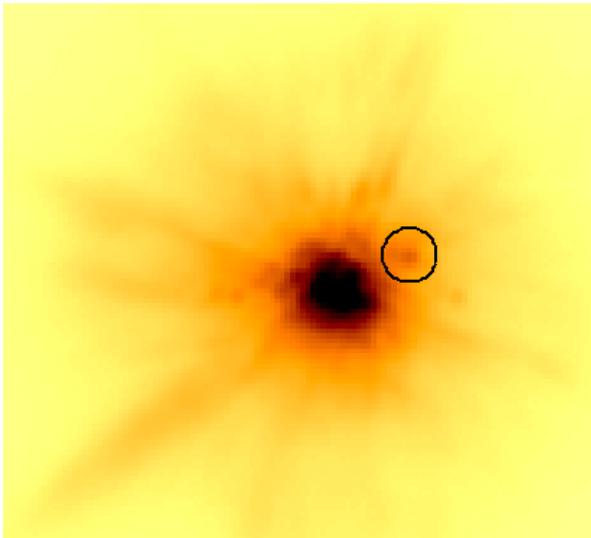}}
  \end{center}
   \caption{I-band image of HD~8673 collected with the AEOS telescope and AO system on 14 October 2004 UT. The companion is circled. The halo of the PSF has many more speckles than the near-IR images, but that is to be expected due to the shorter wavelength of the AEOS images. The image has the same orientation as Figs. \ref{keck_image} and \ref{pharo_image}. This is a sub-image of the full 10\arcsec~image and is roughly 1\farcs8 wide. 
}
 \label{aeos_image}
\end{figure}

Following the Keck II and Hale resolutions, archival observations obtained with the AEOS 3.67 m telescope \citep{vigil1996} were re-examined for a comparable companion detection. On 2004 October 14 UT the star was observed in an effort to detect the bright companion detected by \citet{mason2011}. The observations used the telescope's AO system and the Visible Imager camera \citep{roberts2002}. We collected 1000 frames of the star with the Bessel \textit{I} filter.  After  debiasing, dark subtraction, and flat fielding, the resulting frames were coadded with each frame weighted by its peak pixel to emphasize the best quality data.  The result is shown in Figure \ref{aeos_image}. The companion is harder to see than in the Keck II or Hale images, due to a lower image quality because of the shorter wavelength and because the system has a larger dynamic range in the $I$ filter. 

Again, we measured the astrometry and photometry with the iterative blind deconvolution algorithm, \textit{fitstars}. Photometric error bars were assigned using the technique described in \citet{roberts2005}.    Tables \ref{astrometry_table} and \ref{photometry_table} list the resulting astrometry and photometry measurements.

\subsection{PALOMAR 1.5M OBSERVATIONS}

In 2012 and 2013, we observed HD~8673 four times with the Robo-AO system \citep{baranec2013, baranec2014} on the Palomar 1.5m telescope. On 2012 July 17 UT and  2012 July 18 UT, we used the SDSS \textit{r'}, \textit{i'}, \textit{z'} filters, on  2012 October 6 UT we used the SDSS \textit{i'} \& \textit{g'} filters and on 2013 January 20 UT we used the SDSS \textit{r'}, \textit{i'}, \textit{z'} filters. In 2012, the observations were collected as part of the automated Robo-AO target queue, where the telescope is pointed at the object robotically with a 44\arcsec~wide square field of view. For the 2013 January observations, the star was acquired manually which allowed the star to be positioned at the exact center of camera's field of view, and a 5\farcs5 sub window of the image to be read out at 50.5Hz, which is 5 times faster than the standard rate for the full image. This improves the image quality by reducing the uncompensated intra-exposure stellar image motion. Capturing data in this manner is done infrequently, as it requires human intervention and drastically slows the acquisition rate. 

Each data set was reduced using the standard Robo-AO lucky imaging  pipeline  to perform the image alignment \citep{terziev2013,law2014}.   The pipeline produces four images, with each image selecting either the best 1\%, 10\%, 50\% or 100\% of the images.   Each image was examined visually for companions and no companions were detected.   We computed the dynamic range of the image via the method of \citet{turner2008} and determined that the images have insufficient dynamic range to detect  the companion.  We also examined the wider field of view image taken in 2012 and were able to rule out any additional companions to a limit of 10 magnitudes fainter than the primary.

\section{ANALYSIS}\label{analysis}

\subsection{PROPER MOTION}

Between 2011 and 2012 the measurements from Keck II show the companion moved at a rate of 16.47mas/yr compared to the measured proper motion of HD~8673 of 251.8 mas/yr \citep{vanLeeuwen2007}. The low relative motion of the companion indicates that it shares common proper motion with the primary star and is not a background object. The Keck II data were used to measure the relative proper motion since it is a collection of data taken with the same instrument.  The AEOS and Hale data are also consistent with it being a bound object.

There is very little difference in the short term proper motion ($\alpha$ =237.19$\pm$0.33 mas/yr, $\delta$ = $-$84.64$\pm$0.25 mas/yr, \citealt{vanLeeuwen2007}) and the long term proper motion ($\alpha$ = 237.9$\pm$1.0 mas/yr, $\delta$ = $-$87.2$\pm$1.1 mas/yr, \citealt{hog2000}). Significant differences in these values can indicate the presence of companions \citep{makarov2005}, but their absence is not surprising, especially in the case of a pair with a large differential magnitude and hence large mass ratio. 

\subsection{COMPARISION WITH PRIOR OBSERVATIONS}
 
The two observations of the companion reported by \citet{mason2011} were both optained with visible speckle interferometry using an intensified CCD camera.  The observations were in the Str\"{o}mgren $y$ filter (550$\pm$24nm) and the $\Delta$m from the 2007 measure is estimated as 2.3$\pm$0.5. The quality of the 2001 measure was insufficient for an estimation of the differential magnitude. Using the same speckle interferometry camera as \citet{mason2011}, \citet{hartkopf2009} did not detect the companion in observations using the Mt. Wilson 2.5 m telescope on 2007.8179, only a short time after the 2007.6049 detection in \citet{mason2011}. The observations had a limiting resolution of 0\farcs054.  This would have been sufficient to detect the 0\farcs109 separation detected by \citet{mason2011} measured just a few months earlier.  In addition, no companions were detected by \citet{ginski2012} using the 2.2m Calar Alto telescope and the AstraLux lucky imaging camera in the SDSS \textit{i'} filter on 2011 January 14. 

Our observations using AO have a higher dynamic range than speckle interferometry and would have detected any companions detected with that technique, even if those companions had considerable motion.  The astrometry in Table \ref{astrometry_table} is inconsistent with the  companion reported in \citet{mason2011}. Our non-detection of the candidate companion from \citet{mason2011} and the non-detection in \citet{hartkopf2009} and \citet{ginski2012} lead us to conclude that the \citet{mason2011} detection was spurious. This is not the only known case of false doubles in speckle interferometry; see \citet{mcalister1993} and \citet{tokovinin2012} for examples and further discussion. 

The companion reported in this paper is a redder object than the purported object in \citet{mason2011}. Visible speckle interferometry using intensified CCD cameras can only detect companions with differential magnitudes less than 3.5 \citep{mason1996}. This is too low  to detect the companion reported in this paper and this explains why it was not seen by \citet{hartkopf2009}, or \citet{mason2011}.  The observations of \citet{ginski2012} also did not have sufficient dynamic range to detect the companion.

\subsection{PHOTOMETRIC ANALYSIS}\label{spectral_type}


Based on the distance \citep{vanLeeuwen2007} to HD 8673 AB and the $I$-band apparent magnitude \citep{monet2003}, we computed the absolute magnitude of HD 8673 B in $I$-band of 9.5$\pm$0.94.    Using the 2MASS apparent magnitudes in $J$,$H$, and $K$ \citep{skrutskie2006}, we also computed the absolute magnitudes of companion in the $J$,$H$ and $K$ filters. Our data was taken in  $Jc$, $Hc$ and $Ks$ and we do not have a transformation between these filters into $J$,$H$ and $K$. Instead we use the values from narrow band filters and realize that there is an unestimated error in the magnitudes.  We compared the resulting absolute magnitudes to the values listed on the Dartmouth Stellar Evolution Database\footnote{\url{http://stellar.dartmouth.edu/~models}} \citep{dotter2008}  for a 3 Gyr old star \citep{saffe2005} with a Fe/H of -0.01 \citep{nordstrom2004}.       The $J$,$H$,$Ks$ absolute magnitudes all fall in the same bin.  This bin has a mass of 0.33-0.45 M$_\odot$, $T_{eff}$ of 3520-3690 K, and $\log G$ of 4.94-4.85.  The $I$ absolute magnitude has a much larger error bar and corresponds to a mass range of 0.17-0.44 M$_\odot$.  There is some overlap between the infrared and the $I$ data. This variance in  $I$ band result is not surprising. The $I$ data is the most error prone of all the measurements.  As seen in Figure \ref{aeos_image} the secondary PSF is in the halo of the primary; it also has the largest dynamic range.   
 
\section{DISCUSSION}\label{discussion}

\begin{figure}
\centerline{\includegraphics[width=0.49\textwidth,trim=50 150 50 100,clip]{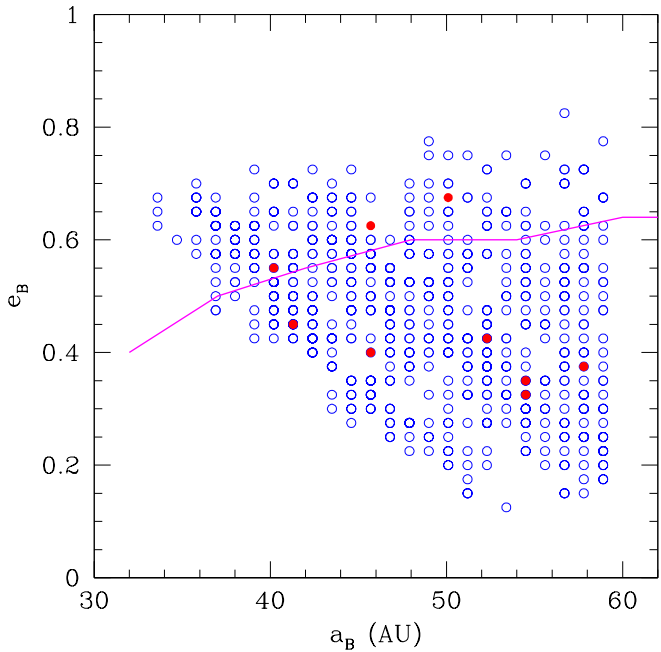}}
\caption{All binary orbits that satisfy the astrometry constraints
  (Table 1) to within $\chi_{\rm reduced}^2 \leq 4$ (blue open
  circles) and $\chi_{\rm reduced}^2 \leq 1$ (red filled circles),
  obtained by scanning through all orbital elements. All these orbits
  are close to edge-on (inclination between $75\deg$ and
  $85\deg$). Only orbits that avoid RV detection are shown (see
  text). Orbits that lie above the magenta lines (for anti-aligned
  apses between the planet and the binary orbits) will destabilize the
  highly eccentric planet, HD 8673b.  
 A non-coplanar situation will also move the magenta lines
  downward.  Admissable binary solutions have semi-major axis $a_B
  \geq 35$ AU. Orbits with very large $a_B$ are disfavoured: 
 }
\label{fig:astrometry-ae}
\end{figure}

While the astrometric data shown in Table 1 does not cover enough of
the orbit to yield a uniqe orbital solution, it does allow us to place
some constraints on the orbital semi-major axis, eccentricity and
inclination. The results are shown in Fig. \ref{fig:astrometry-ae}.
We detail our procedure here, followed by a short discussion on the
implications.

Based on the spectral class of the primary and the secondary, F7V and
 M2V, we obtain a mass sum of $  \sim 1.8 M_\odot$ (masses of $1.4$ and $0.4 M_\odot$ respectively). Adopting this total mass, we then search through the parameter spaces
of all 6 orbital elements to look for acceptable solutions. Due to the
shortness of the arc, there is a wide range of orbital elements that lead to $\chi_{\rm
  reduced}^2 \leq 1$, except for orbital inclinations, which are
strongly constrained to be between   $75$ and $85$ deg from face-on. The binary orbit is nearly edge-on,
and the companion is now lying nearly along our line of sight to the
star. This fact can be simply inferred, without detailed orbital
fitting, from the apparent angular motion of the orbit ($\sim
4\deg/yr$), the apparent separation ($\sim 0.3"$, or $11$AU at a
distance of $36$ pc ), and the system's total mass.

We can pare down possible orbital solutions using two further
constraints.  The first is the lack of evidence for RV residuals from
the binary companion in the Hartmann et al (2010) measurements (over
$6.3$ yrs). The RV residuals, after subtracting the signals from the
highly eccentric planet, is shown to be nearly flat. We adopt a
conservative limit of $\Delta RV \leq 200 m/s$ over $6.3$ yrs, or
$\Delta RV \leq 285 m/s$ over $9$ yrs (the astrometry time-span). The
second, more important, constraint arises from the presence of the
eccentric planet -- the binary orbit ought not to destabilize the
planet ($a_p = 3.02$ AU, $e_p =0.723$). This constraint, as we show
below, rules out a large swath of binary orbits.

Holman \& Wiegert (1999) numerically investigated the stability of a planet orbiting around a star that has a stellar companion. In their simulations, the planet was initially on a circular, coplanar orbit. For a binary mass ratio of $0.28$ (the case here), and a typical binary eccentricity of $0.5$, the binary has to have a semi-major axis $a_B \geq$ $7.1$ $a_p$ for stability. This corresponds to $a_B \geq$ $22$ AU. However, our planet has a high eccentricity, and is possibly not coplanar with the binary orbit. To account for this, we perform our own numerical simulations to define the stability boundary and the results, for coplanar orbits, are also shown in Fig. \ref{fig:astrometry-ae}. These allow us to exclude all orbits with $a_B \leq$ $35$ AU.

While the astrometry data do not exclude orbits that are very wide (e.g., $a_B \geq 60 $AU), these are statistically less probable because the chance of catching the companion at the small separation of $11$ AU is smaller. Roughly, we can peg the most likely binary orbits as $a_B \in [35,60] AU$ and $e_B \leq 0.5$.

This brings us to the question of whether the binary companion has contributed to the planet's high eccentricity. Unless the binary orbit is wide ($a_B \sim 60$ AU) and circular, simulations show the companion strongly influences the planet's orbit. This is even more so if the two orbits are substantially inclined to each other. It is conceivable that HD 8673b was born with a low eccentricity, but is pumped to its current value by the binary companion. It is also conceivable that the high mass of HD 8673b ($M sin i = 14.2 M_J$) is related to this perturbatio -- the large eccentricity swing of the planet may have allowed it to sweep through, and accrete from, a larger area of the protoplanetary disk.

In conclusion, the discovery of the companion of HD 8673 may have provided us with an opportunity to study the impact of binarity on planet formation and migration.  There is still much work to be done on the system.  Additional astrometric observations over the next five to ten years are needed to pin down the orbit of the stellar companion.  It would also be interesting to collect additional RV data to see if there is a radial velocity acceleration caused by the stellar companion, This approach has been fruitful in discovering low mass companions \citep{crepp2012,knutson2014}. An RV acceleration would further constrain the orbit.  An improved orbit  will allow us to quantitatively retrace the dynamical history of the system.

\begin{figure}
\centerline{\includegraphics[width=0.49\textwidth,trim=0 120 0 380,clip]{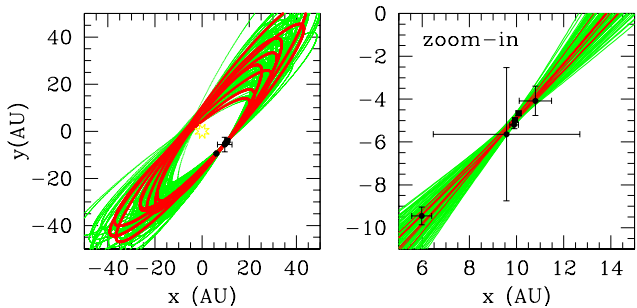}}
\caption{Orbital fit for the acceptable solutions in
  Fig. \ref{fig:astrometry-ae} where red ones satisfy $\chi^2 \leq
  1$, and green ones $1\leq \chi^2 \leq 2$. The binary orbit has to be
  nearly edge-on (with inclination angles between $75$ and $85$
  degrees, in order to satisfy the astrometry observations (points
  with error-bars).}
\label{fig:astrometry-xy}
\end{figure}

\acknowledgments

The research in this paper was carried out at the Jet Propulsion Laboratory, California Institute of Technology, under a contract with the National Aeronautics and Space Administration.  This paper is based on observations from a number of observatories, including the Maui Space Surveillance System, operated by  the US Air Force Research Laboratory's Directed Energy Directorate.  Additional observations were made at the Hale Telescope, Palomar Observatory as part of a continuing collaboration between the California Institute of Technology, NASA/JPL, NOAO, Oxford University, Stony Brook University, and the National Astronomical Observatories of China.  Some of the data  presented herein were obtained at the W.M. Keck Observatory, which is operated as a scientific partnership among the California Institute of Technology, the University of California and the National Aeronautics and Space Administration. The W.M. Keck Observatory was made possible by the generous financial support of the W.M. Keck Foundation. The Robo-AO system is supported by collaborating partner institutions, the California Institute of Technology and the Inter-University Centre for Astronomy and Astrophysics, by the National Science Foundation under Grant Nos. AST-0906060, AST-0960343, and AST-1207891, by a grant from the Mt. Cuba Astronomical Foundation and by a gift from Samuel Oschin. C.B. acknowledges support from the Alfred P. Sloan Foundation. This research made use of the Washington Double Star Catalog maintained at the U.S. Naval Observatory, the SIMBAD database, operated by the CDS in Strasbourg, France, NASA's Astrophysics Data System and  data products from the Two Micron All Sky Survey, which is a joint project of the University of Massachusetts and the Infrared Processing and Analysis Center/California Institute of Technology, funded by the National Aeronautics and Space Administration and the National Science Foundation.

{\it Facilities:} \facility{AEOS (Visible Imager)}, \facility{Hale (PHARO)}, \facility{Keck:II (NIRC2)}, \facility{PO:1.5m (Robo-AO) }





\begin{deluxetable}{cccccccl}
\tablewidth{0pt}
\tablecaption{Photometry\label{photometry_table}}
\tablehead{\colhead{UT} & \multicolumn{6}{c}{Magnitude Difference} & \colhead{Telescope}\\
 &   \colhead{I}    &\colhead{Jc}   &\colhead{Hc}   &\colhead{Fe II} &\colhead{Br$\gamma$}  &\colhead{Ks}   & \\
 &  \colhead{(0.81\micron)} &\colhead{(1.21\micron)} &\colhead{(1.58\micron)} &\colhead{(1.65\micron)}  &\colhead{(2.15\micron)} &\colhead{(2.17\micron)} & }
 \startdata
 2004.7871 &  6.5$\pm1.0$      &...&...&...& ... &  ... & AEOS \\
 2011.5366 &  ...& ...& 4.25$\pm$0.1\phn &... & 4.01$\pm$0.1\phn&... & Keck II\\

 2011.6402 & ...  &...&...&...&   4.50$\pm$0.6\phn & ...  & Hale \\
 2011.8671 &... &4.56$\pm$0.02 &4.20$\pm$0.03 & 4.07$\pm$0.06 & 3.98$\pm$0.08  & ... & Keck II \\
 2012.4837 &... & ...& 4.12$\pm$0.06& 4.03$\pm$0.06 & 3.97$\pm$0.09& ...& Keck II \\
 2013.7426 &  ...&... & ...& ...&...&4.12$\pm$0.06&  Hale 
\enddata
\end{deluxetable}

\end{document}